Efficient formation pathway of methyl formate: the role of OH radicals on ice dust

A. Ishibashi, H. Hidaka, Y. Oba, A. Kouchi, N. Watanabe*
Institute of Low Temperature Science, Hokkaido University
N19W8, Kita-ku, Sapporo, Hokkaido 060-0819 JAPAN.
*e-mail: watanabe@lowtem.hokudai.ac.jp

Abstract

Three structural isomers of the $C_2H_4O_2$ molecule, namely, methyl formate (MF; $HCOOCH_3$), acetic acid (AA; $CH_3COOH$), and glycol aldehyde (GA; $HOCH_2CHO$), have attracted considerable attention as targets for understanding pathways towards molecular complexity in the interstellar medium (ISM). Among these isomers, MF is decisively abundant in various astronomical objects. For various formation pathways of MF, surface reactions on cosmic dust would play an important role. However, when compared to observations, the formation of MF has been found to be relatively inefficient in laboratory experiments in which methanol ($CH_3OH$)-dominant ices were processed by ultraviolet (UV) photons and cosmic-ray analogues. Here, we show experimental results on the effective formation of MF by the photolysis of $CH_3OH$ on water ice at 10 K. We found that the key parameter leading to the efficient formation of MF is the supply of OH radicals by the photolysis of $H_2O$, which significantly differs from $CH_3OH$-rich experimental conditions. Moreover, using an ultra-high-sensitivity surface analysis method, we succeeded in constraining the decisive formation pathway of MF via the photolysis of methoxymethanol (MM; $CH_3OCH_2OH$), which would improve our current understanding of chemical evolution in the ISM.

1. Introduction

Since the first detection by Brown et al. of MF in the star-forming region Sgr B2 (Brown et al. 1975), MF has also been detected in hot corinos (Bottinelli et al. 2007; Öberg et al. 2011) and outflow regions (Arce et al. 2008). Although a plausible formation process of MF in gas-phase is proposed recently (Taquet et al. 2016), the association reactions between radicals on grain surface (Garrod & Herbst 2006) are considered as the important formation process of COMs including MF. In recent years, MF was also found in cold molecular clouds (~10 K) (Bacmann et al. 2012; Jiménez-Serra et al. 2016; Soma et al. 2018). Therefore, several formation mechanisms of MF have been theoretically proposed to explain the presence of abundant gaseous MF in cold molecular clouds: purely gas-phase formation (Balucani et al. 2015), gas-grain model using the chain-reaction mechanism on grain surface (Chang & Herbst 2016), cosmic-ray driven formation (Shingledecker et al. 2018), and non-diffusive formation on grain surface (Jin & Garrod 2020). Those models were successful in explaining the abundance of MF fairly well. However, these are not still conclusive because the detailed information for building reliable models, such as reaction pathway, reaction rate constant, desorption rate by the chemical desorption, is insufficient.

Experimental studies of MF formation in the solid phase under the low temperature conditions have been conducted mainly by the exposure of pure methanol solid to UV photons (Öberg et al. 2009; Paardekooper et al. 2016), because of a good correlation in astronomical observations between MF and $CH_3OH$ (Yang et al. 2021). However, in laboratory experiments involving the photolysis of pure methanol solid, ethylene glycol (EG; $HOCH_2CH_2OH$) and ethanol (EtOH; $CH_3CH_2OH$) were produced in larger amounts than MF (Öberg et al. 2009; Paardekooper et al. 2016; Chuang et al. 2017), suggesting that the following reaction of photoproduced $CH_2OH$ radicals dominates:

$$CH_2OH + CH_2OH \rightarrow HOCH_2CH_2OH (EG) \quad (1),$$
$$CH_2OH + CH_3 \rightarrow CH_3CH_2OH (EtOH) \quad (2).$$

On the one hand, the yields of MF and dimethyl ether (DME; $CH_3OCH_3$) were not significant, and these species would be derived from the following reactions:

$$CH_3O + HCO \rightarrow HCOOCH_3 (MF), \quad (3) \text{ and}$$
$$CH_3O + CH_3 \rightarrow CH_3OCH_3 (DME). \quad (4)$$

In addition, experiments involving a CO-methanol mixture show inefficient formation of MF and were unable to reproduce the observed predominance of MF over other COMs (Chuang et al. 2017).

These previous experiments were performed under $CH_3OH$-dominant conditions, which may be more relevant to the CO freeze-out stage in dark molecular clouds (Cuppen et al. 2017). In ice mantles of grains, methanol would be present in $H_2O$-rich environments ($CH_3OH/H_2O$ ~10-30%, Boogert et al. 2015; Chu et al. 2020, Perotti et al. 2020). Recent experiments simulating the CO-freeze-out stage demonstrated that CO solid cannot fully cover $H_2O$ ice surface. Instead, pure CO solid islands and isolated CO molecules form on water ice in the size range of dust grains (Kouchi et al. 2021a, 2021b). That is, the water ice is always exposed at the surface. On such grain surfaces, CO and its hydrogenated species can be adjacent to $H_2O$ molecules each other. It is very likely that photochemical processes in such $H_2O$-rich sites of the surface differ from those in $H_2O$-poor site on pure CO solid, and thus, there might be unknown processes leading to the effective formation of MF over other COMs. However, the experimental simulation of such processes suffers from technical difficulty in the detection of COM products and their precursors that may form under $H_2O$-rich, $CH_3OH$-poor conditions, where the yields of such species would be most likely smaller than those obtained in pure methanol solid experiments. In the present study, we developed an ultra-high-sensitivity surface analysis method, the so-called $Cs^+$ ion pickup method, which enables us to mass analyse trace adsorbates, including radicals, on an ice surface with sensitivities 2 or 3 orders of magnitude higher than those possible with conventional methods, such as Fourier transform infrared spectroscopy (FTIR).

2. Experiments
2-1. General information about the experimental setup
As shown in Figure 1, the ion pickup experimental apparatus consisted of three parts: a main chamber, an ion source, and a UV light source. The typical pressure of the main chamber during UV irradiation was $10^{-8}$ Pa or less. An Al substrate located in the centre of the main chamber was cooled to 10 K by a He refrigerator and heated to 320 K by an attached ceramic heater. The substrate was electrically insulated from the ground by a sapphire disc and thus could serve as a part of the lens system for ion transport upon application of an electric potential. Approximately 0.3 ML (ML; 1 ML = ~$10^{15}$ molecules $cm^{-2}$) of methanol, as an ice dust analogue, was vacuum-deposited through a capillary plate on ASW of 10 ML (1.5 ML in TPP experiments) pregenerated by background deposition of $H_2O$ vapour on an Al substrate at 10 K. The amount of methanol was estimated by the same manner as our previous experiments (Hidaka et al. 2009). A quadrupole mass spectrometer (QMS) for the ion pickup measurement was installed at an angle of 45° to the substrate surface. At the head of the QMS instrument, the conventional ion source was removed, and instead, ring electrodes divided into two parts for assisting ion transport were installed. This specially designed ring ion lens played a key role in improving the detection sensitivity, which was ~$10^3$ times greater than the sensitivity of a conventional set-up for the ion pickup method. The thickness and bulk composition of the sample solids were monitored by a reflection type FTIR.

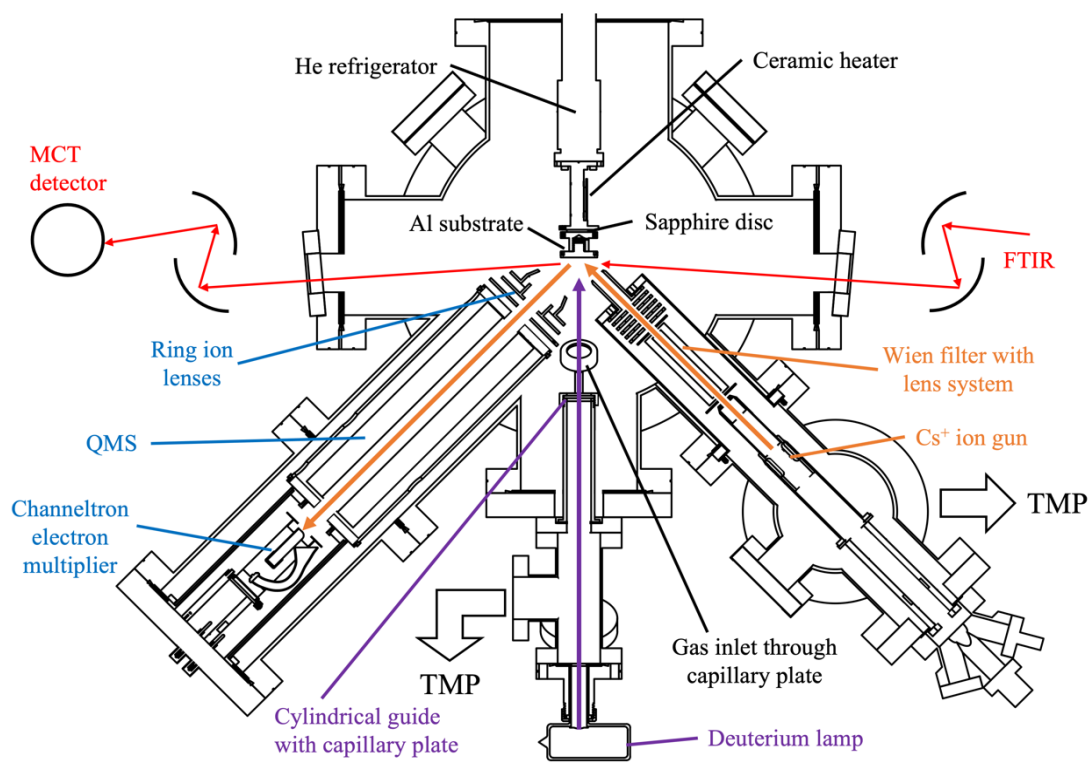

Figure 1 Schematic diagram of the experimental apparatus, orange arrow; Cs+ ion beam, purple arrow; UV light, red arrow; IR beam.

The ion source, a $Cs^+$ ion gun (Kimball Physics, ILG-6), was installed on the mirror surface side of the QMS at an angle of 45° with respect to the substrate surface. A Wein filter system was built in front of the $Cs^+$ ion gun to remove other ion species contained in the $Cs^+$ ion source. The ion source area was differentially pumped to prevent the flow of contamination to the main chamber.

A deuterium lamp (Hamamatsu Photonics: L7293), the wavelength of which was widely distributed across the UV region from 115 to 400 nm, was used as the UV light source. Because the spread of UV light from the lamp may cause emission of photoelectrons and degassing from the wall of the main chamber, the light path was surrounded by a cylindrical guide with a capillary plate on the top and differentially pumped inside. This differential pumping system at the UV light path effectively minimized undesired contamination on the sample surfaces. The average photon flux arriving at the substrate was ~$6 \times 10^{12}$ photons $cm^{-2}$ $s^{-1}$.

2-2. $Cs^+$ ion pickup method

The basic idea of the Cs$^+$ ion pickup method is well-described elsewhere (Kang 2011). In the present experiments, low-energy (~17 eV) Cs$^+$ ions with a flux of ~1 nA picked up the adsorbed molecules (mass number M) on the outermost surface when they were scattered on the substrate. The adsorbates were detected by QMS as Cs$^+$ ion-molecule complexes (mass number 133 + M). The mass number M of the adsorbed species was identified by subtracting the mass number 133 of Cs from the detected mass number (Hang et al. 2001). This method allowed non-destructive detection of adsorbates on the surface, and the number of pickup adsorbates was sufficiently small to have no influence on the surface composition of the sample within the experimental duration. Therefore, continuous measurement was possible in real time (Appendix A1). Our developed method is much more sensitive than conventional methods such as FTIR, as easily confirmed by comparing the detected molecules between the pickup spectrum shown below and the IR absorption spectrum shown in Appendix A2.

Because the electroconductivity of ice is low, the surface of ice can be charged by Cs$^+$ ion injection. It was found that the degree of charging can change the detection efficiency of the ion pickup method. Therefore, the surface potential must be constant through any measurement. We controlled the voltage applied to the Al substrate to keep the surface potential constant. The surface potential, $V_s$, is roughly described by the formula $V_s = V_{Al} + V_{charge}$, where $V_{Al}$ and $V_{charge}$ represent the surface potential applied to the substrate and the potential achieved by charging, respectively. When the degree of charging changes on/in the ice sample, the surface potential can be maintained at a constant value by adjusting $V_{Al}$. The criterion for the adjustment of $V_{Al}$ is to keep the ion pickup signals at a maximum. This procedure is crucial for the experiments because the detection efficiency is extremely sensitive to the configuration of the electric potential between the sample surface and the ion detection systems. In fact, if the $V_{Al}$ is not adjusted with the variation in the surface potential, the pickup signals rapidly disappear.

3. Results and Discussion

Using the Cs$^+$ ion pickup method, we successfully detected a trace amount of photoproduct from methanol on amorphous solid water (ASW) at 10 K. It is worth noting that photoproducts other than H$_2$CO were not detectable by FTIR but were detected by the Cs$^+$ ion pickup method. The results show the efficient formation of MF in consistent with astronomical observations.

Figure 2(a) shows the pickup mass spectrum of methanol adsorbed on ASW at 10 K before UV irradiation. A large peak at Mass 133 is due to the primary Cs$^+$ ion. Peaks at

Mass 151 (=133+18) and 165 (=133+32) originate from $H_2O$ and $CH_3OH$, respectively (hereafter, only the mass of the picked-up species is denoted). For $H_2O$, molecules that are only loosely bonded to the surface are detectable, and thus, the intensities do not reflect the surface number density of the total $H_2O$. The peak at Mass 33 was attributed to $^{13}CH_3OH$ within the methanol reagent with the natural abundance ratio. In addition, peaks derived from simultaneous pickups of multiple $H_2O$ and $CH_3OH$ were detected (see Appendix A3). The spectrum of the sample after UV irradiation is shown in Figure 2(b). The peaks at Mass 30 and 28 correspond to photoproduced $H_2CO$ and CO, respectively. In addition, OH, HCO (or $^{13}CO$) and $CH_3O$ or $CH_2OH$ (or $H_2^{13}CO$) radicals were detected at Mass 17, 29 and 31. Notably, new peaks appeared at Mass 60 and 62, corresponding to $C_2H_4O_2$ and $C_2H_6O_2$, respectively. The small peak observed at Mass 46 can be attributed to HCOOH rather than $CH_3$-derived DME and EtOH, as described in Appendix A4.

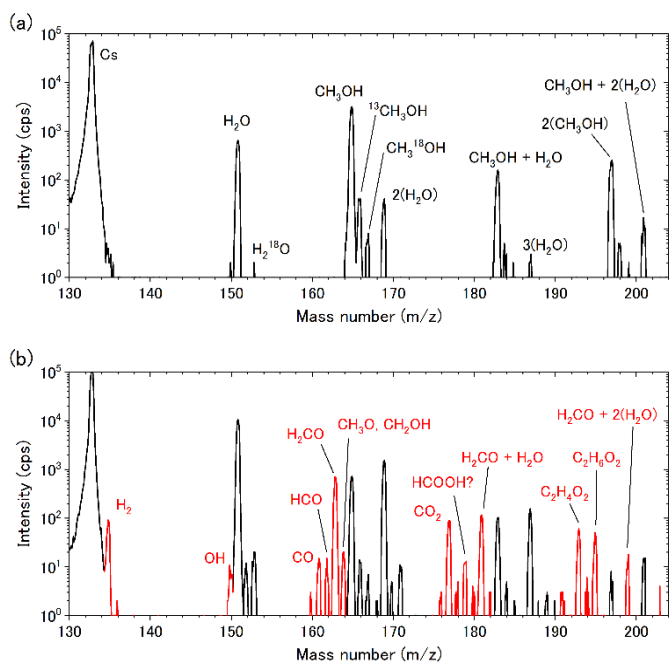

Figure 2 Mass spectra of $CH_3OH$ (0.3 ML) on ASW (10 ML) at 10 K (a) before and (b) after UV irradiation for 3 hours. cps; counts per sec. Peaks newly appeared by UV irradiation are shown in red color.

The peak at Mass 60 ($C_2H_4O_2$) was attributed to GA or MF, and that at Mass 62 was attributed to predominantly EG in previous works (Öberg et al. 2009; Paardekooper et al. 2016; Chuang et al. 2017), but other structural isomers were also possible for each mass: AA at Mass 60 and methoxymethanol (MM; $CH_3OCH_2OH$) and dimethyl peroxide (DMP; $CH_3OOCH_3$) at Mass 62. To identify these molecules, the pickup signals of the

UV products at Mass 60 and 62 were measured while the temperature of the sample increased (hereafter referred to as the temperature-programmed pickup (TPP) spectrum), and the species can be identified by the temperatures at which the signals disappear, which correspond to the desorption temperatures of the target molecules (see Appendix A5). As seen in the TPP spectrum (Figure 3(a)), the signal at Mass 60 drops sharply at ~125 K. As a reference measurement, we obtained the TPP spectra of different samples in which gaseous MF was deposited on ASW (see Figure 3(b)). The disappearance temperature of the peak at Mass 60 for the UV-photolyzed sample was consistent with that of MF for the reference measurement (see Figure 3(c)). As a result, the peak at Mass 60 is assigned to MF. Contributions from other isomers (GA and AA) are negligible since there was little signal above 150 K, which is below the desorption temperature of both molecules (~160 K) (Burke et al. 2015). For Mass 62, there are three main isomers: EG, dimethyl peroxide (DMP; $CH_3OOCH_3$), and methoxy methanol (MM; $CH_3OCH_2OH$). As shown in Figure 3(b), the TPP signal at Mass 62 decreased sharply from ~180 K. The TPP spectra for reference measurements of MM and EG on the Al substrate were obtained (Figure 3(d, e)). The profile of the TPP spectrum for MM matches that for Mass 62 very well, while that for EG does not. Furthermore, the desorption temperature of DMP is known to be lower than that of MM (Zhu et al. 2019). Therefore, we conclude that the photoproduct corresponding to the peak at Mass 62 is mainly MM.

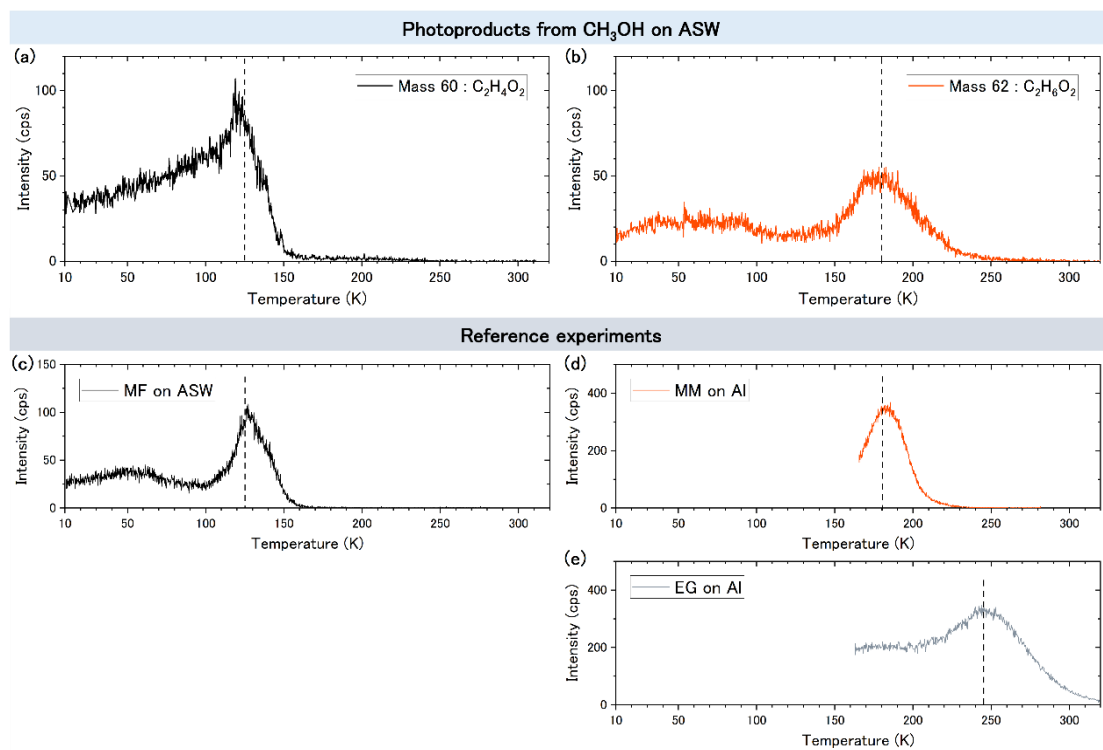

Figure 3 TPP spectra of (a, b) photoproducts at (a) Mass 60 and (b) Mass 62 from CH3OH (0.3 ML) on ASW (1.5 ML) and (c-e) reference experiments for the mass assignments of (c) MF on ASW, (d) MM on Al, (e) EG on Al. The dashed lines indicate the desorption temperature.

In the photolysis of pure methanol solid, the dominant product with a mass of 62 was EG, while MM was not positively identified (Öberg et al. 2009; Paardekooper et al. 2016), which implies that the formation of MM requires water molecules. We also deduce that the efficient formation of MF in the present experiment is correlated with MM formation. To gain more insight into the formation processes, variations in the radicals that may contribute to the formation of MF and MM were measured during UV irradiation of methanol on ASW at 10 K. As seen in Figure 4(a, b), with the consumption of methanol, $CH_3O$ (or $CH_2OH$) radicals appeared first, followed by MM, and finally, MF.

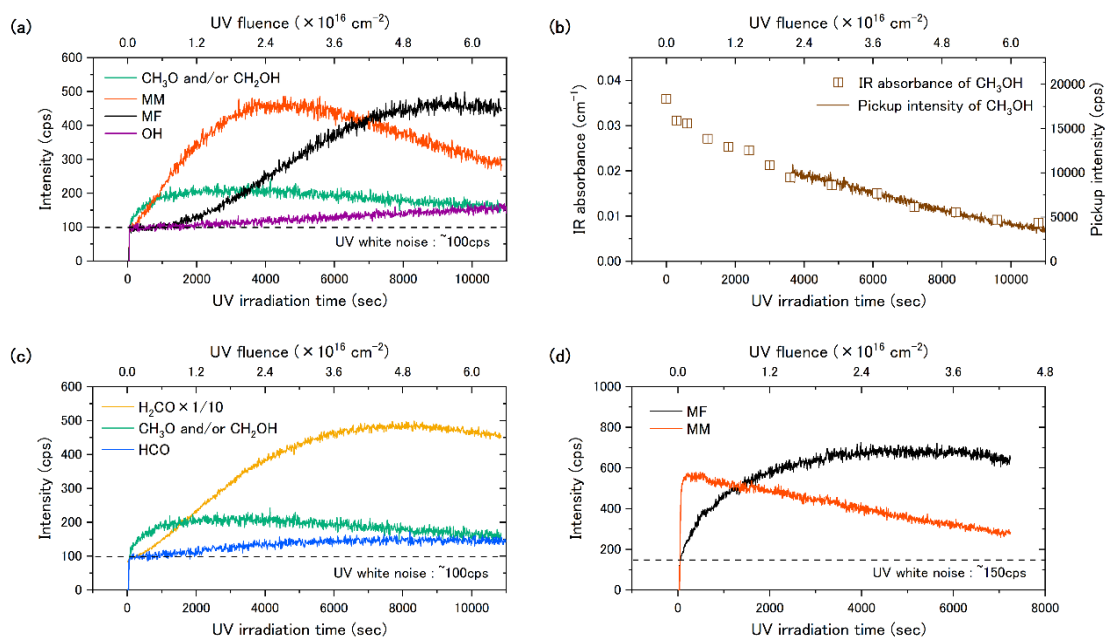

Figure 4 Variations in the signal intensities of surface species generated during UV irradiation at 10 K. (a-c) The sample of CH₃OH (0.3 ML) on ASW (10 ML). (a) Pickup signals of MF, MM, CH₃O and/or CH₂OH, and OH which are photoproducts. (b) Integrated IR absorbance and pickup signals of CH₃OH. The pickup signals before about 3600 s is not shown because of the signal saturation. (c) Pickup signals of CH₃O and/or CH₂OH, HCO, and H₂CO. Since CH₃O and CH₂OH have the same mass number as H₂¹³CO, the value obtained by multiplying H₂CO by the natural abundance ratio was subtracted to obtain the signal of CH₃O and/or CH₂OH only. Similarly, since HCO has the same mass number as ¹³CO, the same treatment was performed. (d) The sample of MM on Al substrate to show that MF is generated from MM. Photons scattered on the substrate collide with the detector, causing UV white noise of ~100 cps for ASW and ~150 cps for Al.

Before discussing MF formation, the formation of MM, as a precursor of MF, is examined. As mentioned above, reactions that occur on the ASW surface should differ from those on the pure methanol solid. EG and MM are inferred to form by reaction (1) and

$$CH_3O + CH_2OH \rightarrow CH_3OCH_2OH(MM) \quad (5),$$

respectively. In other words, the yields of MM and EG depend on the abundance ratio of the CH₃O and CH₂OH radicals on the surfaces. In the case of pure methanol solid, CH₂OH containing photoproducts such as ethylene glycol and ethanol were dominant comparing to the products originated from CH₃O radical (Öberg et al. 2009; Paardekooper et al. 2016). From the latter results, the relationship CH₂OH > CH₃O was deduced. In contrast, the formation of CH₃O leading to MM is expected to be enhanced on ASW. Here, the reactions involved in radical generation on ASW are shown below.

・CH₃O formation (Xu et al. 2007; Shannon et al. 2013; Antiñolo et al. 2016)

$$H_2O + h\nu \rightarrow OH + H \quad (6)$$
$$CH_3OH + OH \rightarrow CH_3O + H_2O \quad (7)$$

・CH₂OH formation (Tachikawa 1993)

$$CH_3OH + h\nu \rightarrow CH_2OH \text{ or } CH_3O \quad (8)$$
$$CH_3O + CH_3OH \rightarrow CH_3OH + CH_2OH \quad (9)$$

In the gas phase, reaction (7) proceeds very fast due to the tunnelling reaction at low temperatures (Xu et al. 2007; Shannon et al. 2013; Antiñolo et al. 2016). On ASW, since H₂O is more abundant than methanol, it is expected that reaction (7) occurs efficiently.

That is, a new CH$_3$O formation pathway in addition to the photo-dissociation of methanol is possible on ASW, and thus, the abundance ratio of CH$_3$O/CH$_2$OH tends to increase. In fact, as seen in Figure 4(a), the OH signal does not appear in the early stages of UV irradiation but appears when MM decreases (that is, MM production stops). This means that OH radicals are rapidly consumed by reaction (7), supporting the above scenario (as a reference, see Appendix A6, for the behaviour of OH on UV-irradiated pure ASW).

CH$_3$O generation can be enhanced by OH radicals, i.e., H$_2$O photodissociation. Because the photodissociation cross sections of H$_2$O and CH$_3$OH in the present UV region are comparable (Cruz-Diaz et al. 2014), the abundance ratio of CH$_3$O to CH$_2$OH can be reflected by the ratio of H$_2$O to CH$_3$OH on the outermost surface of the sample. In fact, the depletion rate of methanol for the photolysis of methanol on the ASW sample was approximately 2 times larger than that for pure methanol solid (Appendix A7), implying the occurrence of reaction (7). In addition, the low detection of EG suggests that CH$_3$O > CH$_2$OH on ice surfaces.

Next, to confirm the pathway from MM to MF, MM deposited on the Al substrate was irradiated by UV photons. As shown in Figure 4(d), MF was found to be efficiently generated by the photoreaction of MM. Reactions leading to MF are thought to be

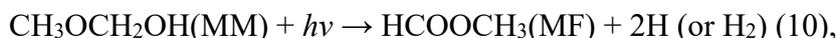
$$CH_3OCH_2OH(MM) + h\nu \rightarrow HCOOCH_3(MF) + 2H \text{ (or } H_2\text{)} \quad (10),$$

together with reaction (3). However, in the present experiment, because HCO is produced via H$_2$CO, its yield is thought to be small. In addition, the behaviours of HCO, H$_2$CO and MF observed in Figure 4 (a) and (c) do not seem to correlate with each other. Therefore, the contribution of HCO to MF formation would be minor under the present conditions.

Chang and Herbst reported that the enrichment of CH$_3$O density on grain surface by reaction (7) enhances MF formation (Chang & Herbst 2016). Although it agrees with our result, our proposed formation mechanism of MF is different from that incorporated in their model. We propose the photodissociation of MM formed by the association of CH$_3$O + CH$_2$OH on grain surface as the additional formation pathway of MF. Besides photolysis and electron bombardments, Chuang et al. experimentally tested MF formation by simultaneous irradiation of CO and CH$_3$OH mixture with UV and H atoms at 14 K (Chuang et al. 2017). They observed the enhancement of relative MF abundance to other major products such as EG and GA when comparing to the pure hydrogenation of the same sample. However, the relative abundance of MF obtained in their experiment was significantly lower than those of astronomical observations (see Figure 7 in Chuang et al. 2017). Consequently, the new reaction pathway proposed in the present study would be the key for the efficient process of MF formation on the cold solid surface.

4. Astrophysical Implications

The present results first show that MF is the overwhelmingly major photoproduct from methanol on ASW while neither EG nor GA are detected. As shown in Figure 5(a), the relative abundance of MF to its precursor, MM, in the present study has a good correlation with observations towards hot corinos of low-mass protostars (Manigand et al. 2020; Jørgensen et al. 2018) and high-mass star-forming region (McGuire et al. 2017 and El-Abd et al. 2019). This correlation would suggest that MF and MM are formed on dust grains in their parent clouds at low temperatures and released into the gas phase during warming up. In addition, the present results indicate that once trace amounts of methanol exist on ice mantles, COMs such as MF can be synthesized without the thermal diffusion of radicals in cold molecular clouds. Because HCO and $CH_3$, both of which are minor products in the present experiment, would be abundant on realistic dust, reactions (3) and (4) are also expected to be promoted, leading to the formation of MF and DME in cold molecular clouds (Figure 5(b)). This inference is consistent with the fact that the $CH_3O$-derived products (MF, DME) were abundantly observed in various astronomical objects compared to the $CH_2OH$-derived products (EG, GA)( Jørgensen et al. 2020; Soma et al. 2018; Herbst & van Dishoeck 2009; Öberg et al. 2009). That is, the number density of $CH_3O$ on ice mantles should be high in cold molecular clouds, and such a situation is most likely brought by reactions (6) and (7).

Occurrence of reactions (6) and (7) requires coexistence of $CH_3OH$ with $H_2O$ on the mantle surface for a long period of time. This situation may not be consistent with the multi-layered ice mantle model, where the $H_2O$ layer will be fully covered with CO and $CH_3OH$ mixed solid as time evolution progressed (Boogert et al. 2015). However, as the ice mantle, a more plausible structure has been recently found (Kouchi et al. 2021a, 2021b). Using the transmission electron microscope observation of ice mantle analogues produced in astrochemically relevant environments, the structure where CO-contained water ice and pure CO solids always coexist on the surface was confirmed.

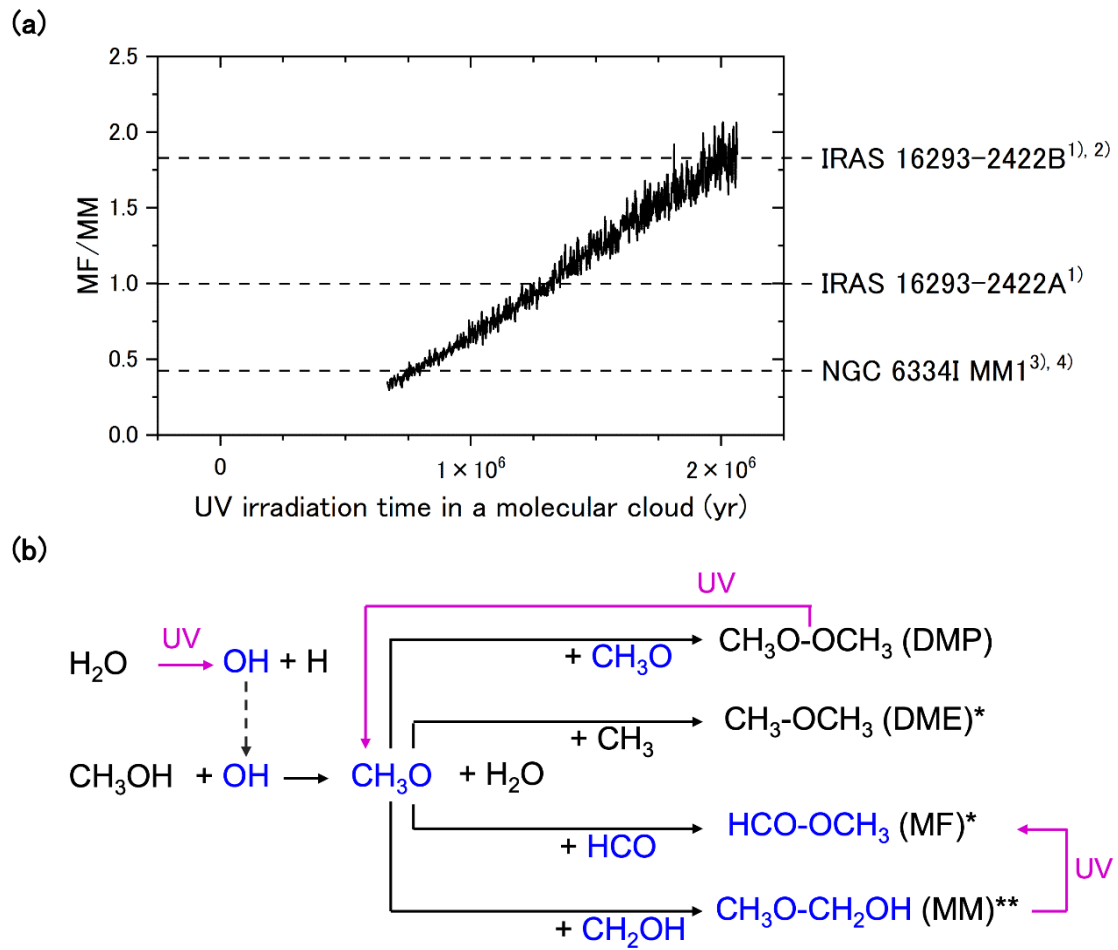

Figure 5 (a) Abundance of MF with respect to MM vs the UV irradiation time corresponding to that in molecular clouds assuming the UV flux of $10^3$ s$^{-1}$ cm$^{-2}$. Here, it is assumed that the pickup efficiencies of MM and MF are comparable. Horizontal dashed line are observational relative abundances. The relative abundance toward IRAS16293-2422A is referred to [1]Manigand et al. (2020). For IRAS16293-2422B and NGC 6344I MM1, the abundances are based on [2]Jørgensen et al. (2018) and [3]El-Abd et al. (2019) for MF and on Manigand et al. (2020) and [4]McGuire et al. (2017) for MM, respectively. (b) Formation path diagram of MF and related COMs formed by radical-radical reactions driven by CH$_3$O on the interstellar dust surface. Blue: Molecules and radicals detected in this experiment. *: Observed COMs in a cold molecular cloud. **: Observed COMs in a high-mass star-forming region.

In this study, we successfully reproduced the observed relative abundances of $C_2H_4O_2$ isomers in cold molecular clouds for the first time. $H_2O$ was found to play an important role in COM formation from methanol. In other words, OH radicals are the key to forming various COMs, especially in cold molecular clouds. Our developed novel apparatus using the $Cs^+$ ion pickup method can be further applied to other related studies, such as nitrogen-bearing COM formation, which is difficult by conventional analytical methods.

This work was supported by JSPS Grant-in-Aid for Specially promoted Research (JP17H06087).

Appendix

A1. Non-destructive $Cs^+$ ion pickup method.

We performed experiments to ensure the non-destructive detection of $CH_3OH$. We monitored the masses of $CH_3OH$ and possible ion-induced fragments (OH, $CH_3O$, and $CH_2OH$) from methanol on ASW without UV irradiation. No change was seen in $CH_3OH$, indicating that the pickup method had little effect on the surface density of $CH_3OH$. The intensities for $CH_3O$ (or $CH_2OH$) and OH were under the detection limit. The intensity for $H_2O$ increased because of the deposition of the residual $H_2O$ gas in the chamber. In the pickup spectrum even after $Cs^+$ ion injection for 3 hours (Figure 6(b)), $CH_3$ (mass 148), MF (mass 193), MM (mass 195), etc., were not detected. Therefore, destruction of the sample by the $Cs^+$ ion beam was almost negligible on the experimental time scale.

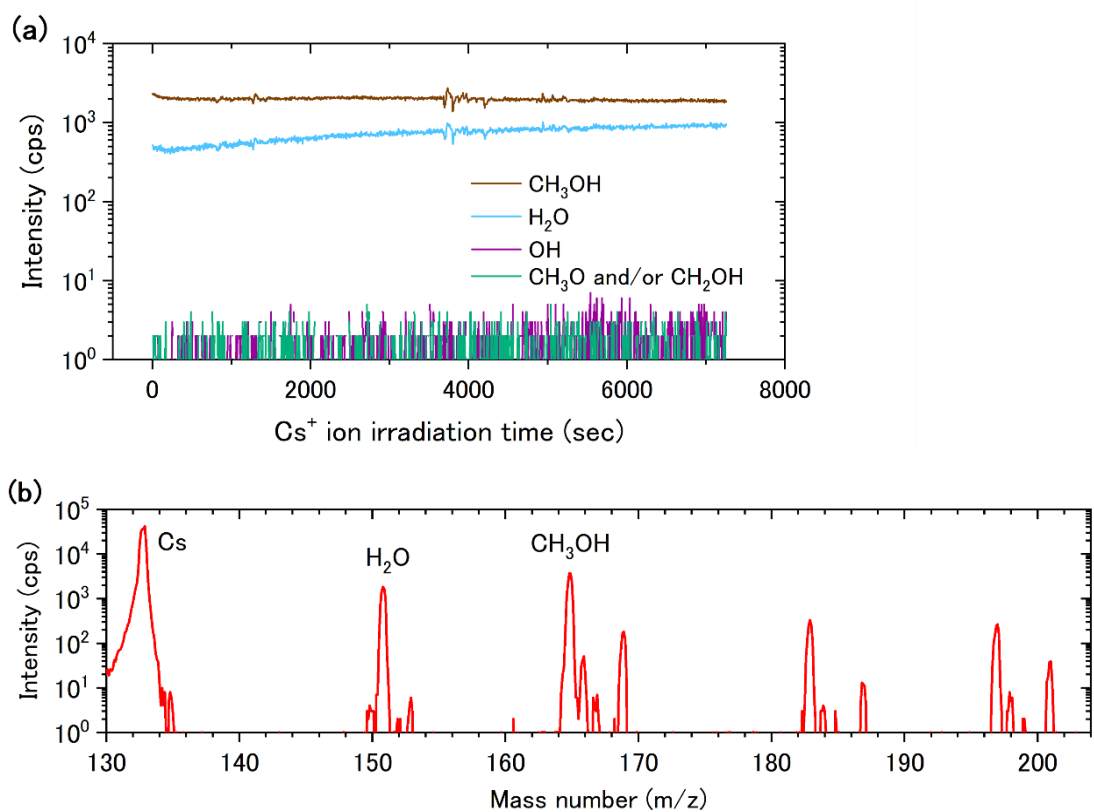

Figure 6. (a) Temporal signal changes in mass corresponding to the masses of CH$_3$OH, OH, H$_2$O and CH$_3$O during Cs$^+$ ion injection to methanol on ASW without UV irradiation. (b) Pickup spectrum after Cs$^+$ ion injection for 3 hours

A2. High sensitivity of the Cs$^+$ ion pickup method.

To demonstrate the high sensitivity of the ion pickup method, we compared the FTIR measurements. Figure 7 shows the IR spectra of UV-photolyzed methanol on ASW under the conditions described in the main text. The amount of methanol that can be detected by the pickup method (Figure 2(b)) at approximately 1000 cps is approximately the detection limit (~0.07 ML) of the FTIR measurements. H$_2$CO, which is produced in comparable amounts to the parent methanol, is expected to be detected at 1723 cm$^{-1}$, but it is hardly detected by FTIR due to its smaller absorption coefficient and overlap with the H$_2$O peak. MF is expected to appear near 1734 cm$^{-1}$ but cannot be detected because its yield is less than approximately 1/10 that of methanol. Accordingly, the pickup method is approximately 100 times more sensitive than FTIR, even for infrared-active species.

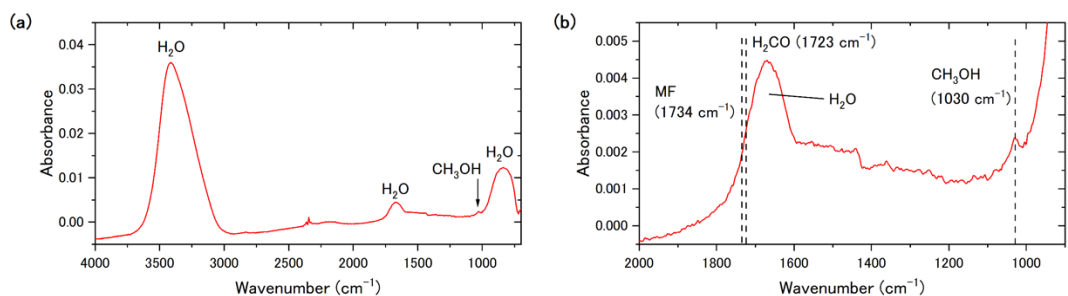

Figure 7. (a) IR spectrum of UV-irradiated methanol (0.3 ML) on ASW (10 ML) for 3 hours, (b) Enlarged view of the 900-2000 cm$^{-1}$ region in (a).

A3. The effect of multiple-molecule pickup in mass spectra

The mass peaks originated by a multiple-molecule pickup sometimes overlap with molecular species synthesized by the chemical reactions; for example, MF and two $H_2CO$ pickup signals appear at the same mass number (M = 60). However, the origin of the mass peak can be easily distinguished in the same manner as molecular identification of the mass number containing isometric molecules described below (A5). In the case of M = 60, the intensities mainly decreased at the desorption temperature of MF but not at the temperature of formaldehyde. It indicates that the peak originates from MF. The mass peaks assigned as a multiple-molecule pickup in Figure 2 were also confirmed to originate from a multiple-molecule pickup by the TPP measurements.

A4. HCOOH (mass of 46) formation from methanol on ASW. No products derived from $CH_3$ were produced.

There are three main structural isomers with a mass of 46. To identify the specific product corresponding to the peak at Mass 46, we obtained the TPP spectrum of this sample in UV-irradiated methanol on ASW (see Figure 8). The TPP intensities did not decrease at the desorption temperatures of DME and EtOH, indicating that both DME and EtOH are minor at the UV-irradiated sample surface. It is presumed that the reaction of the $CH_3$ radical did not occur on ASW, unlike on pure methanol solid. The lack of $CH_4$ production on ASW also supports this hypothesis. This lack of reaction may be due to the low adsorption energy of $CH_3$ on ASW. Instead, the TPP intensities dropped significantly above the desorption temperature of HCOOH. This result clearly shows that the intensity of the peak at Mass 46 is governed by HCOOH. The enhancement of the intensity in the TPP profile above 130 K is thought to originate from the increase in pickup efficiency due to gradual removal of H$_2$O from the solid.

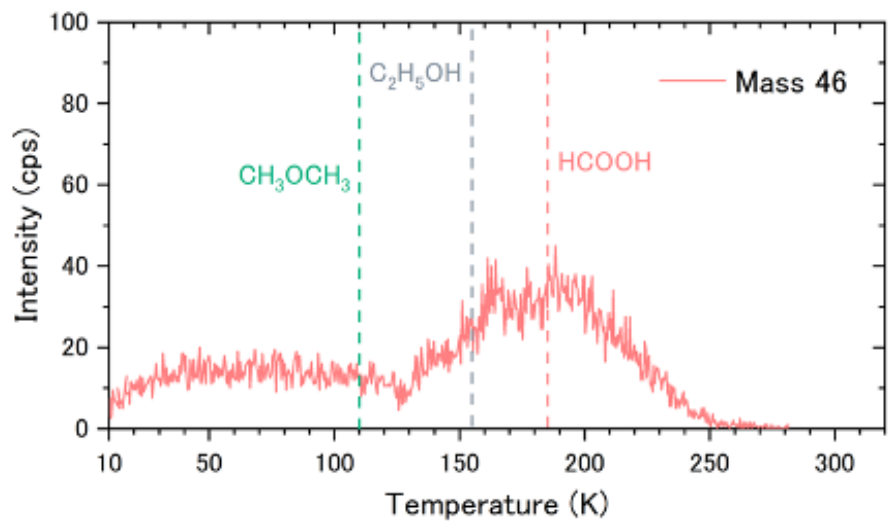

Figure 8. TPP spectrum corresponding to the peak at Mass 46 in UV-irradiated methanol on ASW. The vertical dashed lines indicate the desorption temperatures of DME (~110 K), EtOH (~155 K), and HCOOH (~185 K).

A5. Molecular identification of the mass number containing isometric molecules

Identification of molecular species with the same mass number was performed by measuring the TPP spectrum. In the TPP spectrum, the molecular species can be identified from the differences in the temperatures at which the pickup signals rapidly decrease because each adsorbed molecule has its own desorption temperature. We measured the reference TPP spectra of MF, MM, and EG to obtain each disappearance temperature. Measurements of the reference TPP spectra for MF and ethanol, which have lower desorption temperatures than H$_2$O, were conducted using samples that were prepared by molecular deposition on ASW at 10 K. For EG, which has a higher desorption temperature than H$_2$O, the reference TPP measurement was conducted without a H$_2$O layer sample formed at 165 K. MM was synthesized in the laboratory based on a previously reported protocol (Motiyenko et al. 2018). However, the synthesized MM contained large amounts of contamination (mainly methanol and H$_2$CO). Therefore, the MM sample for the reference TPP measurement was prepared by deposition on an Al substrate at 175 K, where H$_2$CO and CH$_3$OH hardly adsorb.

A6. Behaviour of OH radicals formed by UV irradiation of pure ASW.

When the methanol sample on ASW was continuously photolyzed, OH radicals were expected to form efficiently from the photodissociation of $H_2O$. However, the increase in OH was found to be slow, as shown in Figure 4 (a). We ascribe this finding to the rapid OH consumption by reaction (7), that is, $CH_3OH + OH \rightarrow CH_3O + H_2O$. If this hypothesis is true, the OH intensity should increase immediately when pure ASW is photolyzed. We measured the OH intensities for pure ASW during UV irradiation, as shown in Figure 9. The behaviour of OH supports our hypothesis.

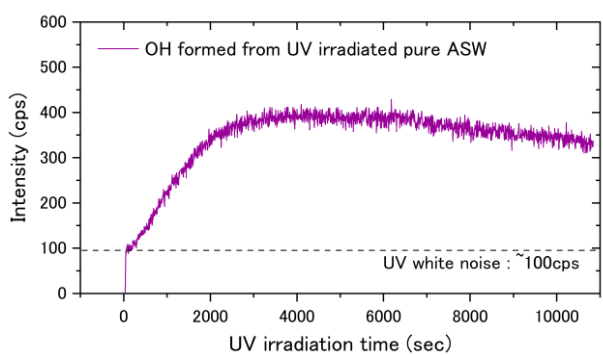

Figure 9. Dependence of the formation of OH radicals from pure ASW with 10 ML at 10 K on the UV irradiation time. In the case of pure ASW, the amount of OH increases rapidly with UV irradiation, which is clearly different from the behaviour of OH detected in the sample of methanol on ASW in Figure 4(a). The UV photons scattered on the substrate and/or wall of the chamber can reach the ion detector, causing UV white noise of ~100 cps.

A7. Enhancement of the methanol attenuation rate by increasing the number of ASW layers.

As discussed in the main text, if reaction (7) enhances the destruction of $CH_3OH$ in addition to direct photodissociation of $CH_3OH$, an increase in OH in the sample should enhance the attenuation rate of $CH_3OH$. Because OH radicals are formed by the photodissociation of $H_2O$, the number of $H_2O$ molecules neighbouring $CH_3OH$ can be used to change the attenuation rate. We consider that the amount of deposited ASW can effectively control the yield of OH near $CH_3OH$. As seen in Figure 10, the attenuation rate of $CH_3OH$ increases as the thickness (amount) of ASW increases. This result clearly shows that $H_2O$ contributes to the enhancement of methanol destruction.

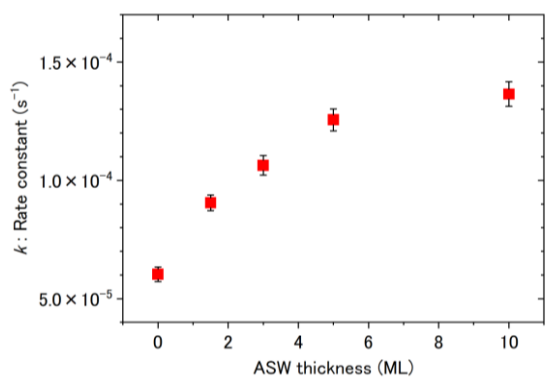

Figure 10. Attenuation rate, $k$, of $CH_3OH$ for the photolysis of methanol (~0.3 ML) on ASW as a function of the ASW thickness. The values of $k$ were obtained by fitting the decrease in $CH_3OH$ absorbance (see Figure 3(b)) to y=A*exp(-$k$*t). The point at ASW of 0 ML corresponds to pure methanol solid (~3 ML). As the ASW film thickness increased, $k$ increased.

A8. Why DMP is not detected despite the $CH_3O$-rich surface
 On the $CH_3O$-rich surface, DMP might be produced by the following reaction.
$$CH_3O + CH_3O \rightarrow CH_3OOCH_3 (DMP)$$
However, O-O single bonds (peroxides) are generally known to be very fragile, perhaps resulting in photo-decomposition into two $CH_3O$ radicals at a high rate (Toth & Johnston 1969). In addition, even if a large amount of $CH_3O$ is produced, it would be converted to $CH_2OH$ by reaction (9).

References
Motiyenko, R. A., Margule`s, L., Despois, D. & Guillemin, J.-C. 2018, PCCP, 20, 5509
Toth, L. M. & Johnston, H. S. 1969, JACS, 91, 1276